\documentclass{aa}
\usepackage{graphicx}
\usepackage{txfonts}
\def\apj{ApJ}%
\def\apjl{ApJ}%
\def\apjs{ApJS}%
\def\ao{Appl.~Opt.}%
\def\aap{A\&A}%

\begin{document}                                                                                                                        
\title{Magnetostatic penumbra models with field-free gaps}
\author{G.B.\ Scharmer
  \inst{1} 
  \and 
  H.C.\ Spruit
  \inst{2} 
  }
\offprints{G.B. Scharmer}

\institute{ Institute for Solar Physics, Royal Swedish Academy of Sciences,
  Alba Nova University Center, 10691 Stockholm, Sweden \\
  \email{scharmer@astro.su.se}
\and 
  Max Planck Institute for Astrophysics, Box 1317, 85741 Garching, Germany \\
  \email{henk@MPA-Garching.MPG.DE}
}
\date{Received (date);  accepted (date)}
                                                                                                                            
                                                                                                                            
                                                                                                                            
\abstract{We present numerical 2D magnetostatic models for sunspot penumbrae consisting of radially aligned field-free gaps in a potential magnetic field, as proposed by Spruit and Scharmer (2006). 
The shape of the gaps and the field configurations around them are computed consistently from the condition of magnetostatic pressure balance between the gap and the magnetic field. The results show that field-free gaps in the {\em inner} penumbra are cusp-shaped and bounded by a magnetic field inclined by about $70^\circ$ from the vertical. Here, the magnetic component has a Wilson depression on the order $200$--$300$~km relative to the top of the field-free gap; the gaps should thus appear as noticeably elevated features. This structure explains the large variations in field strength in the inner penumbra inferred from magnetograms and two-component inversions, and the varying appearance of the inner penumbra with viewing angle. In the outer penumbra, on the other hand, the gaps are flat-topped with a horizontal magnetic field above the middle of the gap. The magnetic field has large inclination variations horizontally, but only small fluctuations in field strength, in agreement with observations.}
                                                                                                                            
\maketitle
                                                                                                                            
\section{Introduction}
The spatial resolution achieved with adaptive optics systems, implemented recently on major solar telescopes (Rimmele 2000, Scharmer et al. 2000, Soltau et al. 2002), now approaches the values needed to resolve the intrinsic length scales expected for magnetic, as well as nonmagnetic, structures on the solar surface. The spectacular convergence between numerical simulations (Carlsson et al. 2004, Keller et al. 2004, Steiner 2005) and recent observations of the small scale magnetic field (Lites et al. 2004) provides confidence, both in our theoretical understanding of these structures and in the power of the best current observations to test the theory convincingly. 

The more puzzling phenomenology of sunspots still awaits a similar breakthrough. The discovery that penumbral filaments have dark cores flanked by lateral brightenings (Scharmer et al. 2002), however, strongly suggests that the fundamental scales of penumbral filaments are now being resolved as well. These recent observations have also highlighted the evolutionary connection of penumbral filaments with umbral dots and light bridges, as well as their morphological similarities. This hints at the possibility of a common underlying structure. 

The center-to-limb variation of  structure in sunspot images provides some geometrical information on vertical structure (e.g. Lites et al. 2004), but  most information about the 3rd dimension is encoded in spectral line profiles and their polarization properties.  Inversions of these data into a vertical structure model, based on techniques developed by e.g. \ Skumanich and Lites (1987), Ruiz Cobo and del Toro Iniesta (1992) and Frutiger et al. (1999), are very far from unique and must be regularized by assumptions about the vertical structure of the thermodynamics and magnetic field configuration in the atmosphere. Alternatively, forward modeling by radiative transfer in detailed structure models is used (e.g. \  Solanki \& Montavon 1993, Martinez Pillet 2000,  M{\"u}ller  et al. 2002).

On the theoretical side, 3D MHD simulations including full radiative transfer (Stein  \& Nordlund 1998, V\"ogler et al. 2003) are able to model increasingly large volumes of the solar atmosphere.  Recent progress in understanding of umbral dots with such simulations (Sch\"ussler \& V\"ogler 2006) raises hopes that convergence between theory and observation of sunspot structure is a realistic prospect for the near future. 

Much of the current thinking about penumbral structure and Evershed flows is based on 1D MHD simulations of thin magnetic flux tubes assumed to move in a background of different magnetic properties (Schlichenmaier et al. 1998a, b). These simulations have led to the view that the magnetic field in the observable layers of the penumbra is intrinsically far removed from its lowest energy state, the potential field. Interpretations of polarized spectra obtained at low spatial resolution made in a similar view are the so-called embedded flux tube or `uncombed' penumbra models (Solanki \& Montavon 1993).

Spruit and Scharmer (2006, hereafter SS06) have questioned these interpretations of penumbra fine structure. In the layers above the photosphere, the magnetic field is already so dominant that deviations from a potential field configuration is unlikely on the observed time scales of penumbral structure. The Alfv\'en crossing time on which a magnetic structure changes,  unless it is close to a potential or at least a force free field, would only be of the order of seconds for the embedded tubes of width $\sim 100$ km assumed in the models. It seems unlikely that such embeddings will be found in realistic 3D MHD simulations, such as may become possible in the near future.

In SS06 an alternative scenario is proposed for understanding penumbra fine structure, its magnetic field configuration and its energy balance. It assumes the existence of field-free, radially aligned gaps below the visible surface, intruding into a nearly potential field above. In this model, the dark penumbral cores outline the centers of the field-free gaps, analogous to the dark lanes running along light bridges in spots (Berger \& Berdyugina 2003, Lites et al. 2004), and on a smaller scale the `canals' seen in strong field regions outside spots (Scharmer et al. 2002). Remarkably, the dark cores of light bridges have already been reproduced in full 3-D radiative MHD simulations (Nordlund 2005, Heinemann 2006). They are due to the enhanced opacity associated with the higher gas pressure in the field-free gaps combined with an overall drop of temperature with height. Outlining the `cusps' of these gaps, they appear as elevated above and dark relative to their surroundings. This is consistent also with the appearance of the dark cores of penumbral filaments.

Another important motivation for embedded tube models has been the small length scale on which the inclinations of field lines vary. Vertical length scales on the order of a hundred km above the photosphere, as inferred from observations (Sanchez Almeida  \& Lites 1992, Solanki et al. 1993) led to the idea that the penumbral atmosphere could not be the smooth structure expected from a potential field. The assumption of non-potential structure in the form of small-scale inclusions within the height of formation of a spectral line may have seemed straightforward. In SS06 we have shown that this seemingly obvious conclusion is nevertheless erroneous. Not only are small scale variations consistent {\em with}, they are actually an inevitable property {\em of} potential fields. By the nature of the Laplace equation, an inhomogeneity of wavelength $\lambda$ imposed at the boundary of the domain decays into it with an (e-folding) length scale $\lambda/2\pi$. For horizontal structure with $\lambda\sim 1"$ in the penumbra, this predicts a vertical length scale of $\sim$100 km. 

In this way the interpretation of filaments as gaps, solving the penumbral heat flux problem, and the observation of small scale field inclination variations mutually support each other. It does so within the elegance of a potential field model which is expected, on theoretical grounds, to hold approximately in the atmosphere.

The shape of a gap (i.e., its width as a function of depth) is determined by the condition of  pressure balance between the surrounding magnetized fluid and the nonmagnetic stratification in the gap. This is the same mathematical problem as finding the shape of the sunspot flux bundle from the balance of pressures at its outer boundary (e.g., Jahn and Schmidt 1994).

The analytic potential field model used in SS06 took this force balance into account only in the coarsest sense. More quantitative detail is needed when comparing the model with observations. For example, the top of the gap must have a `cuspy' shape in order for the explanation of dark cores (see above) to hold, and the observed ranges of field line inclinations in inner and outer penumbra must be reproduced. In this paper we address this with numerical solutions for the field configuration, assuming  a pressure stratification in the gap approximating that of the normal convection zone, and a similar but reduced pressure distribution inside the magnetic field. These assumptions lead to models that are distinctly different for the inner and outer penumbra and that allow observed properties of dark-cored filaments to be explained.

\section{Embedded flux tubes models}

In this section we present some critical thoughts on recent embedded  flux tube models.

The uncombed penumbra model (Solanki \& Montavon 1993), often also referred to as the embedded flux tube model, was developed in order to explain the strong vertical gradients in the inclination of the magnetic field inferred from Stokes~V spectra (e.g., Sanchez Almeida \& Lites 1992), while avoiding strong curvature forces (Solanki et al. 1993). As explained above, potential fields avoid these problems automatically, removing much of this motivation. Nevertheless, the line of reasoning was straightforward and made contact with existing `magnetoconvection' views of the penumbra in which vertical displacements of approximately horizontal field lines played a major role. 

Solanki and Montavon (1993) proposed that these penumbral flux tubes could be modeled as flux tubes with circular cross sections, internal field lines aligned with the flux tube and external field lines wrapping around the flux tube. As pointed out in SS06, such round flux tubes cannot be in magnetostatic equilibrium. The surrounding magnetic field of such a flux tube must vanish at the top and bottom. At the sides of the flux tube, the surrounding magnetic field strength will be increased by the presence of the flux tube. This produces forces that will compress the flux tube horizontally and make it expand upwards at its top and downwards at its bottom. In SS06 we estimated that this will flatten the flux tube in tens of seconds. This is due to the low density and correspondingly high Alfv\'en speed in the line-forming layers. Unsuccessful efforts to construct reasonable magnetostatic flux tube models (Borrero 2004), high-light the fundamental difficulties of such embeddings in the penumbra atmosphere and furthermore demonstrate that these problems are in no way reduced for {\em thin} flux tubes. The suggestion that magnetostatic equilibrium may be achievable with partly flattened flux tubes (Borrero et al. 2006b), remains a speculation.

Our objection to round flux tubes  therefore applies also to the moving tube model of Schlichenmaier, implemented in a thin flux tube approximation. At a sufficiently large depth below the surface where the gas density is high enough, the flattening time scale (Alfv\'en crossing time) may be large,  but the mismatch of time scales will become a problem already long before a tube reaches the penumbral photosphere.

Long, stable flux tubes, maintaining their identity and extending all the way from the inner to the outer penumbra as in Schlichenmaier's moving tube model provide a possible explanation to the Evershed flow, but their assumed long-lived identity also constitutes a hindrance to explaining penumbral heating. 

With the degrees of freedom in existing embedded tube models, they can explain observed polarized spectra (e.g. Bellot Rubio et al. 2004, Borrero et al. 2004, 2005, 2006a,b, Martinez Pillet 2000). While these inversions represent a significant advance in exploiting spectropolarimetric  information, confidence that they substantiate flux tube models is misplaced, since inversion of line profiles is fundamentally non-unique. While it can rule out classes of models, it can not be used as positive evidence for a model that fits the data. At a basic level, the very nature of radiative transfer, with broad contribution functions at each wavelength, also makes it very difficult to distinguish discontinuities from smoother variations on the basis of observed (polarized) spectra. 

These flux tube models, while meant to represent a structure embedded in a surrounding medium, do this in a physically inconsistent way. The radiative transfer model is a single ray intersecting  a constant-property flux tube, while the background atmosphere is assumed similarly homogeneous. The displacement of field lines needed to accommodate the structure is ignored. An embedding would displace the surrounding magnetic field lines (unless violation of div ${\bf B}=0$ is assumed), causing inhomogeneity of order unity around it in field strength, field line directions, or both. The agreement with observations obtained with such models is thus of unquantifiable significance.

While not necessary for interpreting the observed line profiles, attempts are sometimes made to fit the results into a concept of nearly horizontal tubes extending from the inner penumbra to its outer edge (e.g. Bellot Rubio et al. 2004). The magnetic field of such flux tubes would have to be almost exactly parallel to the $\tau=1$ surface of the penumbra, else they would quickly run out of the line forming region (cf. Schlichenmaier \& Schmidt 2000, Bellot Rubio et al. 2004).  This is in disagreement with measured field line inclinations (e.g.  Bellot Rubio et al. 2004, Borrero et al. 2005, Langhans et al. 2006). Attempts to make these measurements agree with the notion of long horizontal tubes are unconvincing.

We emphasize that two-component inversions clearly provide indisputable evidence for the existence of large inclination and field strength gradients in the penumbra and thereby crucial information not attainable by other means. 
However, such inversions do not allow firm conclusions about the underlying structure responsible for these gradients.

\section{Periodic potential field with field-free gaps} 

Following SS06, we develop our 2D potential field model in cartesian coordinates. The $z$-coordinate is the vertical direction, $y$ the horizontal direction parallel to the filament (also called here the {\em radial coordinate}), and $x$ the horizontal direction perpendicular to the filament  (also called the {\em azimuthal} coordinate). The structure is assumed independent of the $y$-coordinate. This is justified by the fact that penumbral filaments are long compared to their widths. Whereas a 3D model obviously would be more satisfactory, the present relatively simple model is adequate for demonstrating major differences between `gappy' magnetic fields in the inner and outer penumbra and for comparing these models with observations.
 
We assume that there are no field lines entering or leaving the gap, i.e., the discontinuity  follows field lines, and that the magnetic field is a potential field outside the gap and identically zero inside the gap. Because of the periodic field and symmetry assumed (shown in Fig.~1), we can expand $B_x$ as a sine series and $B_z$ as a cosine series

\begin{equation}
\label{eq:bx0}
B_x = \sum\limits_{n=1}^{\infty} f_n(z) \sin (k_n x) 
\end{equation}
and
\begin{equation}
\label{eq:bz0}
B_z = \frac{g_0}{2} + \sum\limits_{n=1}^{\infty} g_n(z) \cos (k_n x) ,
\end{equation}
where $g_0$ constitutes a height independent vertical field component equal to the average vertical flux density,
\begin{equation}
k_n = n \pi /L
\end{equation}
and $L=S/2$ equals half the separation between two filaments.
Since the discontinuity is aligned with a field line, the magnetic field is 
divergence-free ($\mathbf{\nabla \cdot B} = 0 $) everywhere, which implies that 
\begin{equation}
\label{eq:fngn}
f_n = - \frac{1}{k_n} \frac{{\rm d}g_n}{{\rm d}z} .
\end{equation}
The assumption that the magnetic field is zero inside the gap and a potential field 
outside the gap implies that the height dependent sine and cosine coefficients can be 
obtained for $n=0,1,2...$ as
\begin{equation}
\label{eq:fn1}
f_n(z) = \frac{2}{L} \int\limits_{0}^{L} B_x \sin (k_n x)~{\rm d}x
=\frac{2}{L} \int\limits_{x_{\rm g}}^{L} \frac{\partial{\phi}}{\partial{x}} 
\sin (k_n x)~{\rm d}x 
\end{equation}
and 
\begin{equation}
\label{eq:gn1}
g_n(z) = \frac{2}{L} \int\limits_{0}^{L} B_z \cos (k_n x)~{\rm d}x
=\frac{2}{L} \int\limits_{x_{\rm g}}^{L} \frac{\partial{\phi}}{\partial{z}}
\cos (k_n x)~{\rm d}x ,
\end{equation}
where $x_{\rm g}=x_{\rm g}(z)$ outlines the current sheet constituting the interface between the 
field-free and magnetic atmospheres.
Our goal is to derive a second relation between $f_n$ and $g_n$. Integrating the first
equation by parts and using that $\sin(k_n L) = \sin (n \pi) = 0$, we obtain
\begin{equation}
\label{eq:fn2}
f_n = -\frac{2}{L} \phi (x_{\rm g},z) \sin(k_n x_{\rm g}) - \frac{2 k_n}{L} \int\limits_{x_{\rm g}}^{L} 
\phi \cos (k_n x)~{\rm d}x .
\end{equation}
To evaluate Eq. (6), we use that
\begin{eqnarray}
\frac{\rm d}{{\rm d}z} \left( \int\limits_{x_{\rm g}}^{L} \phi \cos (k_n x)~{\rm d}x \right) \nonumber\\
  = \int\limits_{x_{\rm g}}^{L} \frac{\partial{\phi}}{\partial{z}} \cos (k_n x)~{\rm d}x 
 - \frac{{\rm d}x_{\rm g}}{{\rm d}z} \phi (x_{\rm g},z) \cos(k_n x_{\rm g}) ,
\end{eqnarray}
giving
\begin{equation}
g_n = \frac{2}{L} \frac{\rm d}{{\rm d}z} \left(\int\limits_{x_{\rm g}}^{L} \phi \cos (k_n x)~{\rm d}x \right)
+ \frac{2}{L} \frac{{\rm d}x_{\rm g}}{{\rm d}z} \phi (x_{\rm g},z) \cos(k_n x_{\rm g}) 
\end{equation}
Comparing to Eq. (\ref{eq:fn2}), we obtain after simplifications
\begin{equation}
g_n = - \frac{1}{k_n} \frac{{\rm d}f_n}{{\rm d}z} - \frac{2}{L k_n} \frac{{\rm d} \phi(x_{\rm g},z)}{{\rm d}z} \sin (k_n x_{\rm g}) .
\end{equation}
We introduce the variable $D_{\rm g}$ 
\begin{equation}
D_{\rm g}  = \frac{{\rm d} \phi(x_{\rm g},z)}{{\rm d}z} ,
\end{equation}
which can be evaluated as
\begin{equation}
\label{eq:btz1}
D_{\rm g} = \frac{{\rm d}x_{\rm g}}{{\rm d}z} \frac{\partial{\phi}}{\partial{x}} +
\frac{\partial{\phi}}{\partial{z}} = \frac{{\rm d}x_{\rm g}}{{\rm d}z} B_{{\rm g}x}(x_{\rm g},z) + B_{{\rm g}z}(x_{\rm g},z)  .
\end{equation}
Using that the field is tangent to the interface
\begin{equation}
 \frac{B_{{\rm g}x}}{B_{{\rm g}z}} = \frac{{\rm d}x_{\rm g}}{{\rm d}z}
\end{equation} 
to eliminate $B_{{\rm g}x}$, we obtain
\begin{equation}
D_{\rm g} = B_{{\rm g}z} \left[1 + \left(\frac{{\rm d}x_{\rm g}}{{\rm d}z}\right)^2  \right] ,
\end{equation}
where $B_{{\rm g}z}=B_z(x_{\rm g},z)$ is the vertical magnetic field component at the boundary.
Finally, using Eq. (\ref{eq:fngn}) we obtain
\begin{equation}
\label{eq:gn2}
g_n = \frac{1}{k_n^2}\frac{{\rm d}^2g_n}{{\rm d}z^2} - \left[1 + \left(\frac{{\rm d}x_{\rm g}}{{\rm d}z}\right)^2 \right]
\frac{2}{L k_n} \sin(k_n x_{\rm g}) B_{{\rm g}z} .
\end{equation}
This equation can be written as
\begin{equation}
\label{eq:gn3}
g_n = \frac{1}{k_n^2}\frac{{\rm d}^2g_n}{{\rm d}z^2} + \left[ 1 + \left(\frac{{\rm d}x_{\rm g}}{{\rm d}z}\right)^2 \right] s_n B_{{\rm g}z} ,
\end{equation}
where
\begin{equation}
s_n = - \frac{2}{L k_n} \sin(k_n x_{\rm g}) = \frac{2}{L} \int\limits_{x_{\rm g}}^{L} \cos (k_n x)~{\rm d}x .
\end{equation}
Comparing to Eq. (\ref{eq:gn1}) shows that the $s_n$ can be identified with the cosine
coefficients of a vertical magnetic field of unit amplitude that has no horizontal 
gradients within $x_{\rm g}<x<2L-x_{\rm g}$ and that is zero outside this interval. 
We also note that the last term vanishes when $x_{\rm g}(z)=0$. This implies that 
the magnetic field is potential for all $x$, which is the case above the height where the 
field-free gap closes. For such heights, the equations for all $g_n$ are uncoupled 
\begin{equation}
g_n = \frac{1}{k_n^2}\frac{{\rm d}^2g_n}{{\rm d}z^2} 
\end{equation}
and the solutions are (eliminating solutions that grow exponentially with height): 
\begin{equation}
\label{eq:gn5}
g_n(z) = g_n(z_0) \exp(-k_n (z-z_0)) ,
\end{equation}
where $z_0$ is the height above which the gap is closed for all $z$.

At heights where the field-free gap is open, Eq. (\ref{eq:gn2}) shows that the equations 
for all $g_n$ are {\it coupled} through the $B_{{\rm g}z}$ term. This term can be expressed as a 
weighted sum of all $g_n$ terms, see below. A direct solution of Eq. 
(\ref{eq:gn2}) would
therefore correspond to a relatively large matrix equation with  $M N$ unknowns, where $M$ is the number of depth points and $N$ the number of cosine coefficients used to expand $B_z$ in the $x$-direction.

\section{Numerical solution}


We introduce a depth grid $(z_1,z_2,....z_M)$, where $z_1$ corresponds to the lower boundary and $z_M$ to the upper boundary, identified with the {\it first} depth point for which the gap is closed. $\Delta z$ is the grid spacing, and  we represent derivatives at depth point $m$ numerically as

\begin{equation}
\label{eq:d2gn}
\frac{{\rm d}^2g_n(z_m)}{{\rm d}z^2} = \frac{1}{\Delta z^2} (g_n(z_{m-1}) - 2g_n(z_m) +g_n(z_{m+1}))  .
\end{equation}
With the boundary condition discussed below and assuming that the shape of the gap is
given (this will be determined by force balance across the discontinuity, see Sect. 5), Eq. (\ref{eq:gn3}) can be written as a 
matrix equation for each $g_n$,
\begin{equation}
\label{eq:an}
\mathbf{A_n \cdot g_n = S_n \cdot B_{{\rm g}z}} ,
\end{equation}
where boldface quantities are either vectors or matrices and $\mathbf{S_n}$ is a diagonal
matrix. This can be inverted to express $g_n$ in terms of $B_{{\rm g}z}$

\begin{equation}
\label{eq:gn4}
\mathbf{g_n = A_n^{-1} \cdot S_n \cdot B_{{\rm g}z}} .
\end{equation}
$B_{gz}$ can be expressed in terms of $g_n$ as
\begin{equation}
\label{eq:btz2}
B_{{\rm g}z} = g_0 + 2 \sum\limits_{n=1}^{N} g_n(z) \cos (k_n x_{\rm g}(z)) ,
\end{equation}
where $g_0$ is assumed given.
Note that at heights where the gap is {\it open}, this equation contains a multiplicative 
factor of two compared to what is 
expected from  Eq. (\ref{eq:bz0}), because that equation gives a {\it boundary} value that is the average
of the values at both sides of the discontinuity. 
This equation can be represented as a matrix operation 
\begin{equation}
\label{eq:btz3}
\mathbf{B_{{\rm g}z} = 2 \sum\limits_{n=1}^{N} C_n \cdot g_n + d +g_0} ,
\end{equation}
where $\mathbf{d}$ represents the lower boundary condition. 
Combining this equation with Eq. (\ref{eq:gn4}), we obtain a matrix equation for $B_{{\rm g}z}$,
\begin{equation}
\label{eq:btz4}
\mathbf{B_{{\rm g}z} - 2 \sum\limits_{n=1}^{N} ( C_n \cdot A_n^{-1} \cdot S_n) \cdot B_{{\rm g}z}
= d + g0} .
\end{equation}
This shows that we can build up a matrix equation for $B_{{\rm g}z}$, i.e. involving only
the vertical component of the magnetic field {\it along the discontinuity}. Having thus calculated
$B_{{\rm g}z}$, the solution for each $g_n$ can be obtained by solving Eq. (\ref{eq:gn4}) for 
each $n$ separately.

We note that the matrix equation for $B_{{\rm g}z}$ allows $x_{\rm g}$ to be arbritarily chosen and
need not be at discrete grid points. When combined with the requirement of force balance across the discontinuity (Sect. 5), the equation thus defines a smooth solution $x_{\rm g}(z)$.

The lower boundary condition for $g_n$ is easily expressed in terms of a given 
vertical magnetic field $B_z$ at the lower boundary. 
A reasonable assumption is that the vertical
magnetic field is constant, $B_z = B_0$ for $x_{\rm g}(z_1) < x < 2 L - x_{\rm g}(z_1) $, 
implying that $g_n = B_0 s_n$ and thereby also fixing the constant value of $g_0$ 
at the lower boundary. The effects of this approximation can be reduced by increasing the depth of the lower boundary.

\section{Magnetostatic potential field model}
Because of the potential field assumed, both the field-free and magnetic components of the
atmosphere are in hydrostatic equilibrium but with a gas pressure that is in general different
at any height for the two components. 

Equilibrium across the gap dictates that the sum of 
gas pressure and magnetic pressure must be continuous across the gap. With a
given gas pressure variation with height in the two components, also the 
variation of the field strength along the discontinuity is given. To satisfy that constraint
with an assumed given magnetic field at the lower boundary,
the {\it shape} of the discontinuity, i.e. the variation of $x_{\rm g}$ with $z$, must adjust 
itself to produce the field strength needed to comply with force balance across the 
discontinuity. This is a free boundary problem, first applied to sunspot models by Schmidt and
Wegmann (1983) and later by Jahn and Schmidt (1994). To solve this problem in the context of the
present model, we  first write
\begin{equation}
\label{eq:pf}
P_{\rm f} = P_{\rm m} + B_{\rm g}^2/(2\mu_0).\label{peq}
\end{equation}
where $P_{\rm f}$ and $P_{\rm m}$ are the gas pressures in the field-free and magnetic components
respectively and $B_{\rm g}$ is the magnetic field strength along the discontinuity. In addition to its  $(x,z)$-components, i.e., in the plane perpendicular to the filament, the magnetic field has a component $B_y$ parallel to it. This component is  assumed homogeneous, except in the gap, where it vanishes.

To specify the problem, the gas pressures  $P_{\rm m}$ and $P_{\rm f}$ have to be given as functions of depth $z$. Since the gap communicates directly with the convection zone surrounding the spot, its pressure can be approximated from a model for the mean pressure stratification in the convection zone. The gas pressure in the magnetic field is more uncertain. An important measure for $P_{\rm m}(z)$ is the {\em Wilson depression}, the depth below the normal solar surface of the optical depth unity surface, which, however, is known with some accuracy only for the umbra. The choice of $P_{\rm m}(z)$ also influences the height $z_0$ where the gap closes. To complete the model definition, we have assumed that $z_0=0$ everywhere in the penumbra, that is, the top of the gaps is at the level of the normal solar photosphere. This agrees with the appearance of the bright filaments, in particular its dependence on viewing angle and disk position, but must be considered an assumption subject to future improvements. With this assumption, the Wilson depression $\delta z_{\rm W}$ of the {\it magnetic} component
becomes a part of the solution of the problem. In the inner penumbra, we shall find a value of about $300$ km, a plausible value in view of the observed value in the umbra, $\delta z_{\rm Wumbra}\approx 400$~km.  A similar value ($300$~km) was also found for the height of a dark cored light bridge by Lites et al. (2004), based on purely geometrical arguments.

The magnetic field strength along the boundary is calculated as
\begin{equation}
\label{eq:bg}
B_{\rm g}^2 = B_y^2+B_{{\rm g}x}^2+B_{{\rm g}z}^2=B_y^2+B_{{\rm g}z}^2 \left[1 + \left(\frac{{\rm d}x_{\rm g}}{{\rm d}z}\right)^2\right] .
\end{equation}
All quantities, here and in the following, refer to conditions along the discontinuity.
Combining  Eqs.  (\ref{eq:pf}) and (\ref{eq:bg}), we can write
\begin{equation}
E(z)=0 ,
\end{equation}
where
\begin{equation}
E(z)=B_{{\rm g}z} \left[ 1 + \left(\frac{{\rm d}x_{\rm g}}{{\rm d}z}\right)^2 \right]^{1/2}-(2 \mu_0(P_{\rm f}-P_{\rm m})-B_y^2)^{1/2} .
\end{equation}
To find the shape of the gap, we have chosen to minimize the integral of $E^2$ along the discontinuity, $L$, with respect to $x_{\rm g}(z)$. Thus $L$ is given by 
\begin{equation}
L  = \int\limits_{0}^{s_{\rm max}} E(s)^2 {\rm d}s = \int\limits_{z_{\rm min}}^{z_{\rm max}} E(z)^2 \left[1 + \left(\frac{{\rm d}x_{\rm g}}{{\rm d}z}\right)^2\right]^{1/2} {\rm d}z ,
\end{equation}
where $s$ is a coordinate along the discontinuity and ${\rm d}s=({\rm d}x^2+{\rm d}z^2)^{1/2}$. To achieve this, $x_{\rm g}(z)$ was defined at a small number of nodes and interpolated between the nodes by cubic splines, following Schmidt and Wegmann (1983) and Jahn and Schmidt (1994). $E(z)$ was linearized with respect to small perturbations in $x_{\rm g}$ at the nodes and the linearized equation solved with least squares methods. The solutions converged in $5$--$10$ iterations to errors in $E(z)$ of less than about $1$--$3$~mT.

As an approximation to the field-free gas pressure, $P_{\rm f}(z)$, we have taken a polytropic stratification,  i.e., a scale height that varies linearly with $z$, such that $H=H_{f1}=390$ km at $z=-500$ km and $H=H_{f2}=160$ km at $z=0$.  Over the relevant depth range this is a fair match to a mean solar model.
For the magnetic atmosphere we assumed that the pressure variation with $z$ was identical to that of the field-free atmosphere, but scaled by a constant $C$. The implied temperature variation with height is thus identical for the two components.

The assumption that the gap closes at $z=0$ together with the assumed known pressure variations $P_{\rm f}$ and $P_{\rm m}$ means that the scale factor $C$ is determined by the field strength at the height where the gap closes. Here, the azimuthal field $B_x$ must vanish for symmetry reasons, and force balance across the discontinuity requires that at height $z=0$,
\begin{equation}
P_{\rm m}(0) = P_{\rm f}(0) - (B_y^2 + B_{{\rm g}z}(0)^2)/(2 \mu_0) .
\end{equation} 
The vertical field component, $B_{{\rm g}z}(0)$, depends on the average vertical field, $\bar B_z$, and the shape of the gap. For a gap with a flat top, $B_{{\rm g}z}(0)$ is close to zero, whereas for a pronounced cusp shape, $B_{{\rm g}z}(0)$ is closer to $\bar B_z$. The assumption that the gap closes at $z=0$ therefore  implies a relation between the gas pressure in the magnetic component, the radial field $B_y$, the average vertical field $\bar B_z$ above the surface, and the {\it shape} of the gap.

The model used for the gas pressure in the field-free component is such that the magnetic atmosphere is completely evacuated at the top of the gap (at $z=0$) when the field strength is $170$~mT at that height. For stronger magnetic fields, the gap must close {\it below} the height $z=0$. This means that the top of the gap will be associated with a Wilson depression relative to the quiet sun, but does not imply that it will be invisible since the gas above it has reduced gas pressure, and thereby low opacity. 

\subsection{Model parameters and properties}
In Table I are shown parameters of four models for the magnetic field discussed in the
following.   These parameters are related to $\bar B_z$ and $B_y$ through
\begin{equation}
\bar B_z = \bar B \cos(\bar \gamma)
\end{equation}
and
\begin{equation}
B_y = \bar B \sin(\bar \gamma) ,
\end{equation} 
where $\bar B$ is the average field strength and $\bar \gamma$ is the average inclination.
We have chosen separations $S$ between the filaments that are in the range 
$500$--$1000$~km, in rough agreement with those found for dark-cored filaments (Langhans 2006). The average magnetic fields and inclinations used are similar to those found by Borrero et al. (2005), discussed in Sect. 6.2.  
\begin{table}[tbh]
  \centering
  \begin{tabular}{llllll}
    \hline
    \hline
    \vspace{1mm}
    Case & $\bar B$ (T) & S (km) & $\bar \gamma (^\circ)$ \\
    \hline
    I & 0.10 & 1000 & 75  \\
    II & 0.14 & 1000 & 60  \\
    III & 0.18 & 1000 & 45  \\
    IV & 0.18 & \phantom{0}500 & 45  \\
    \hline\hline
  \end{tabular}
  \caption{Model parameters. $\bar B$ is the average field strength, $S$ the separation between the filaments and $\bar \gamma$ the average inclination of the magnetic field.}
  \label{tab:cases}
\end{table}
Figures (1)--(3) show the results of these calculations. Figure 1 shows the shape of the discontinuity and the field lines for the calculations made. Also shown as a dashed horizontal line is the height at which the gas pressure in the magnetic component equals the gas pressure in the field-free component at $z=0$. In the absence of radiative transfer calculations, this is used as a proxy for the continuum forming layer, referred to in the following as the penumbral photosphere, and therefore also as an indication of the Wilson depression. In Figs. (2) and (3) are shown the inclination angle and field strength variations along this photosphere.

{\it Case I} corresponds roughly to conditions in the outer penumbra. The average magnetic field chosen is strongly inclined (average inclination $75^\circ$ with respect to the vertical) and weak (average field strength $100$~mT), the separation between two gaps, $S$, was set at  $1000$ km. Figure 1 shows the shape of the gap and the field lines calculated. We note that the discontinuity is flat-topped over more than  $400$ km above the center of the field-free gap. Figure 3 shows the variation of the field strength as function of $x$ at the photosphere (full), $100$ km (short dashes) and $200$km (long dashes) resp. above the photosphere. The field strength above the center of the gap is identical to that of the radial ($B_y$) component, showing that $B_z$ vanishes above the gap. Figure 2 shows the variation of the magnetic field inclination, calculated as $\tan^{-1} (B_y/B_z)$, with $x$ at $z=0$ (full) and at $z=100$ km (dashed). The inclination varies from $90^\circ$ above the gap to $54^\circ$ midways between two gaps. Thus the magnetic field is associated with large inclination variations but small variations in field strength. The Wilson depression in the magnetic component is about $60$ km.
\begin{figure}[htbp]
 \centering \includegraphics[bb=55 112 740 404, clip, width=1.00\hsize]{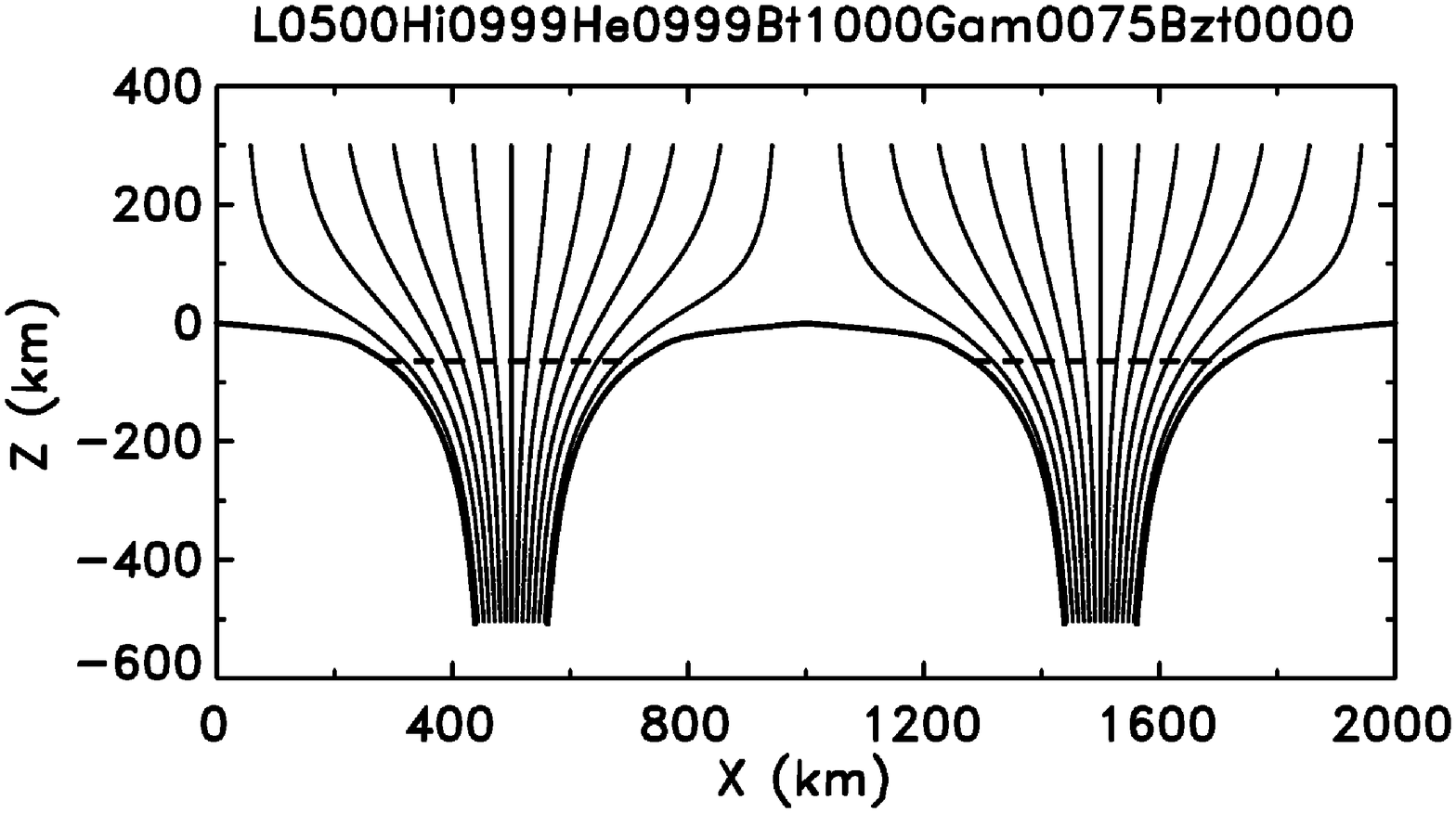}
 \centering \includegraphics[bb=55 112 740 404, clip, width=1.00\hsize]{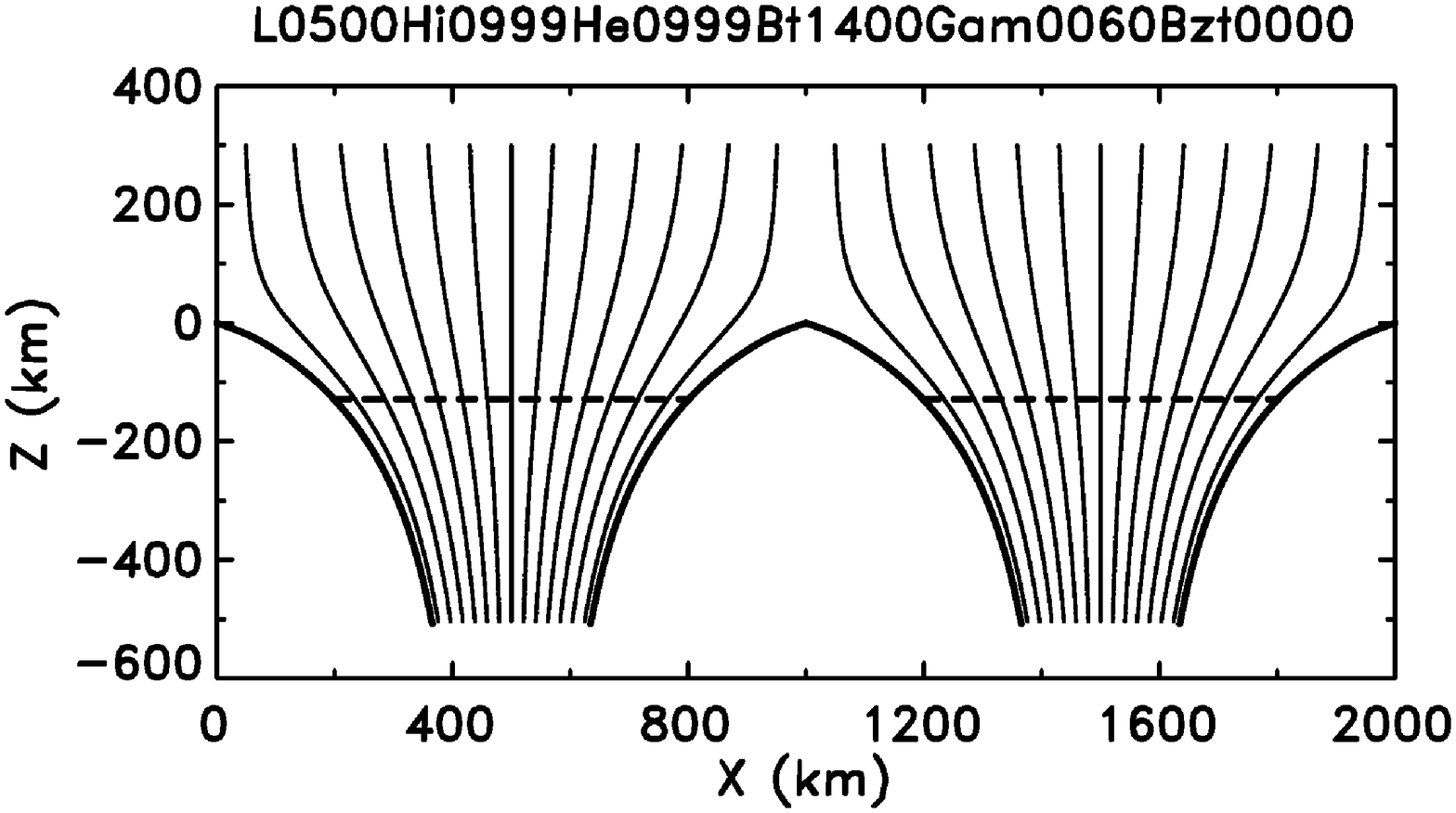}
 \centering \includegraphics[bb=55 112 740 404, clip, width=1.00\hsize]{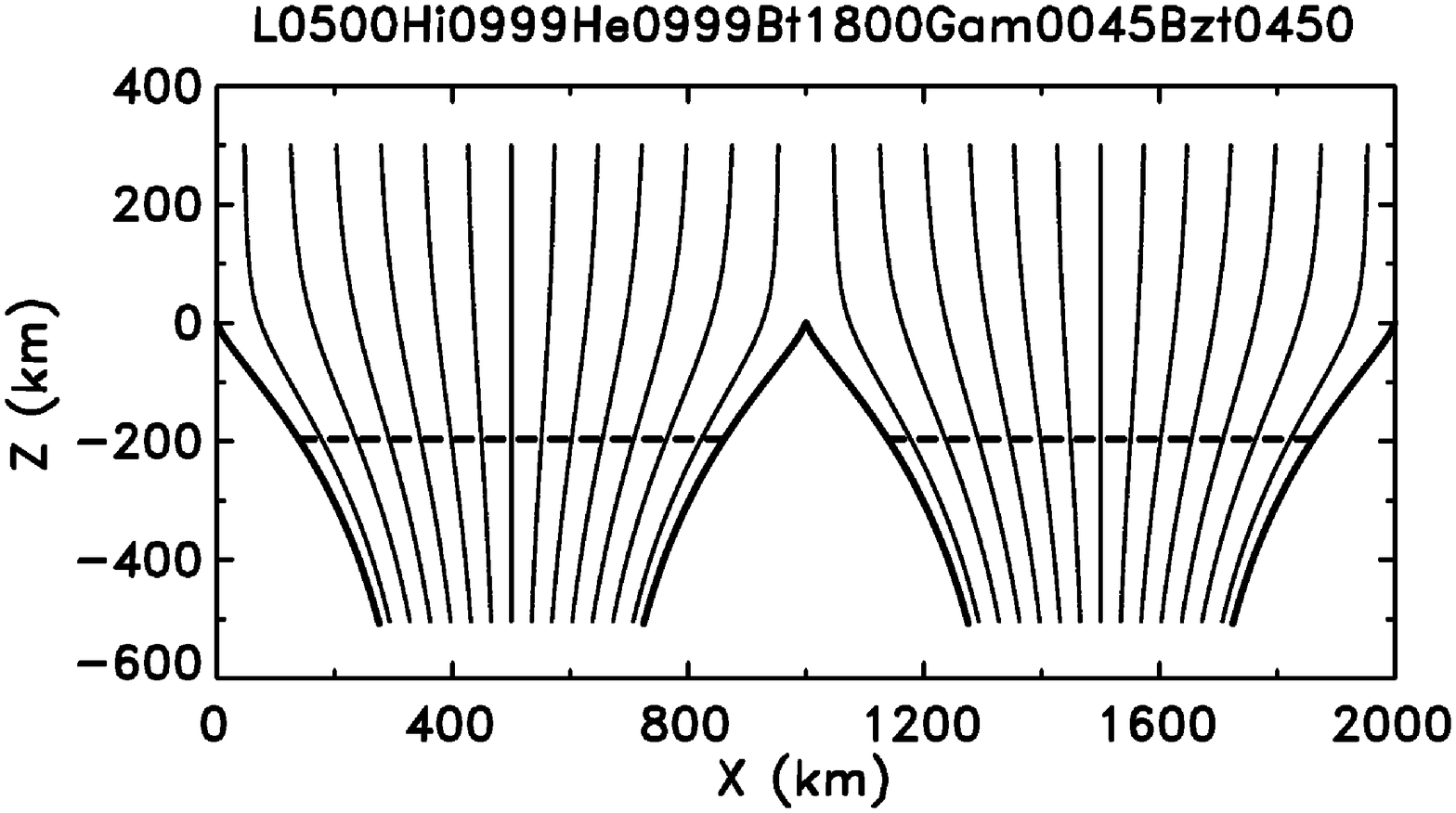}
 \centering \includegraphics[bb=55 68 740 404, clip, width=1.00\hsize]{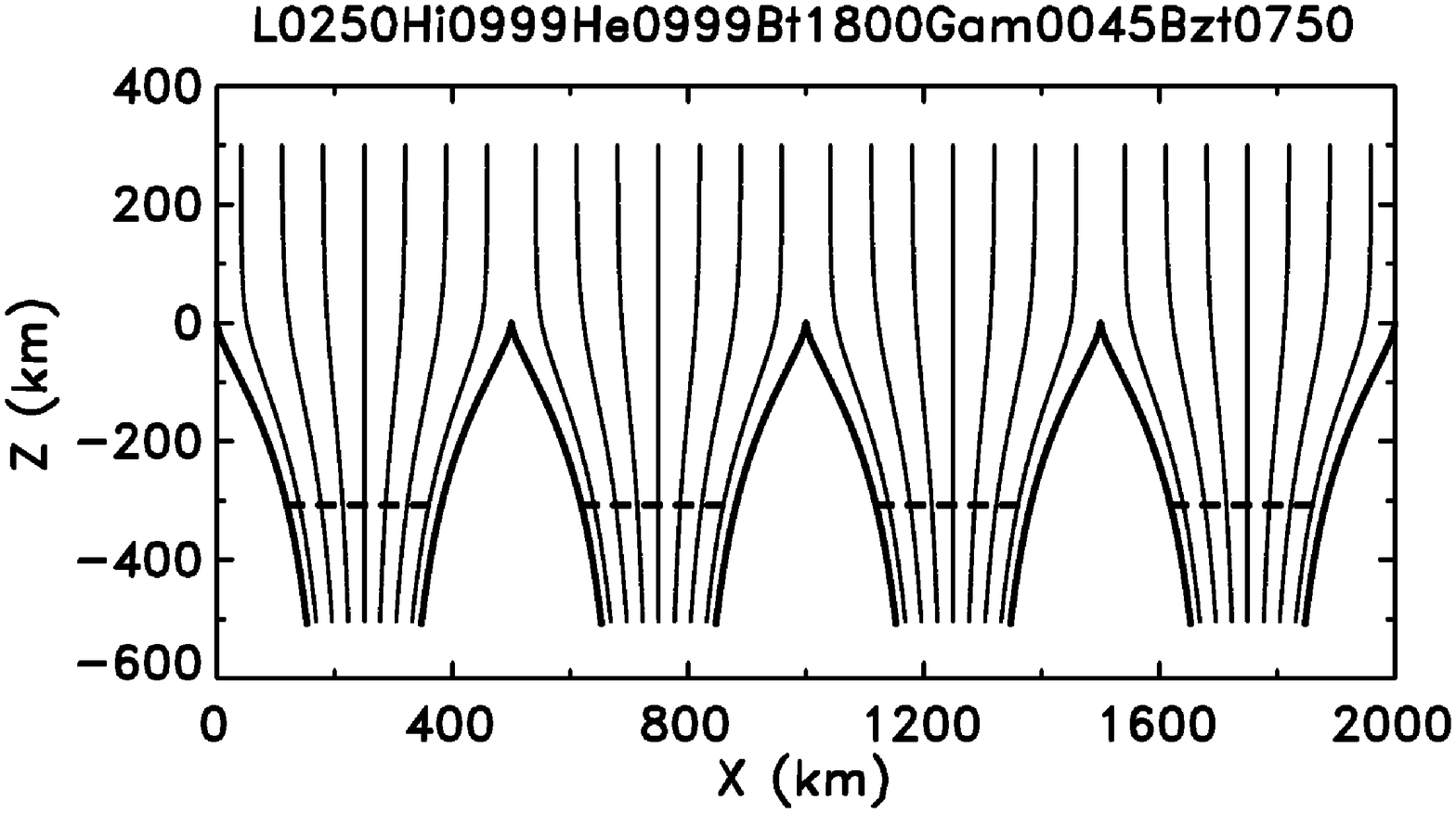}
     \caption{\small Gap shape (thick lines) and field lines for Case I (top), representing the 
      outer penumbra, Case II representing the mid penumbra and Cases III-IV the inner
      penumbra.
      The dashed horizontal line indicates the height where the gas pressure in the 
      magnetic component is equal to that of the field-free component at $z=0$.
             }
     \label{cores1}
\end{figure}
\begin{figure}[htbp]
 \centering \includegraphics[bb=55 112 740 404, clip, width=1.00\hsize]{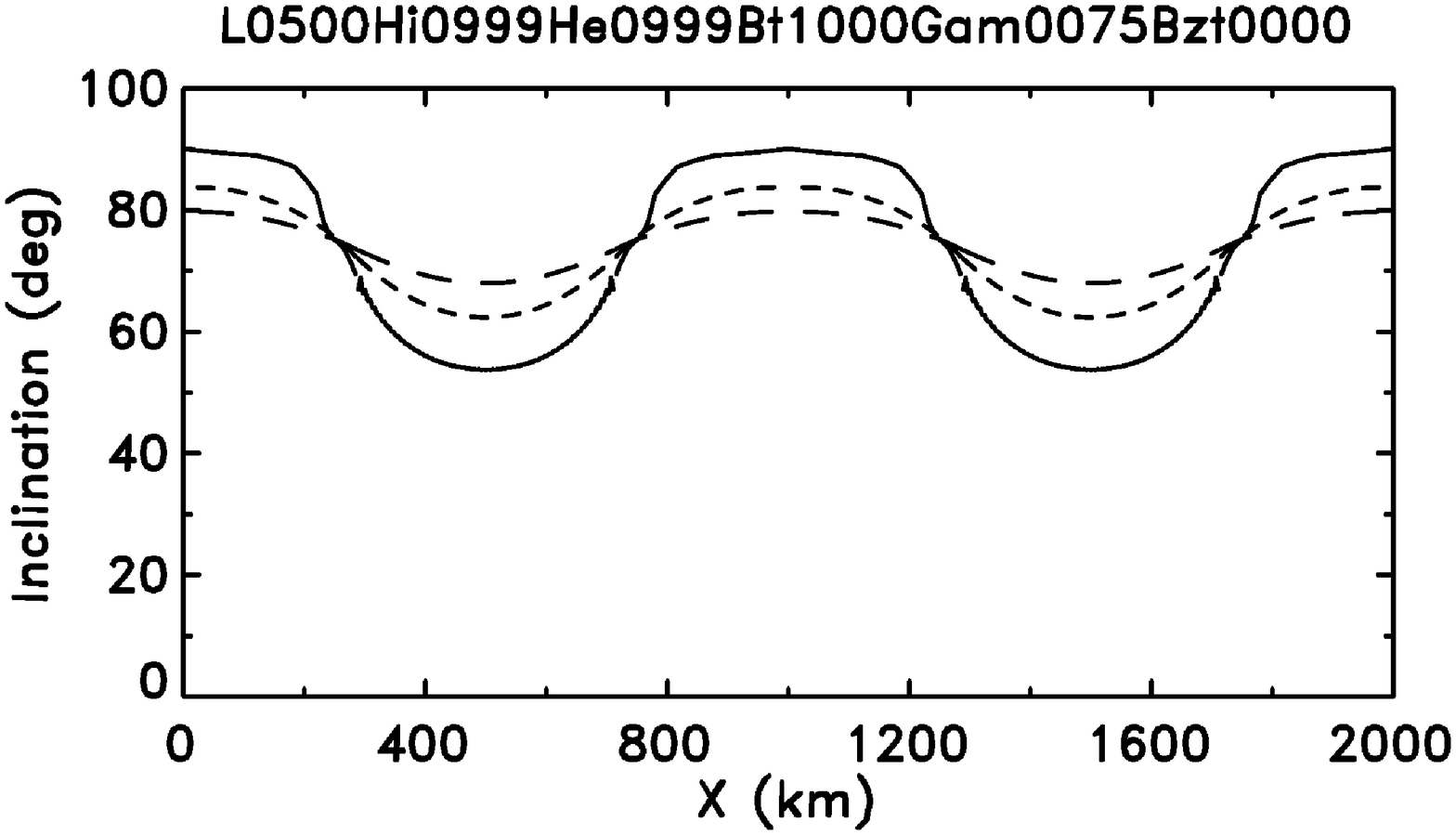}
 \centering \includegraphics[bb=55 112 740 404, clip, width=1.00\hsize]{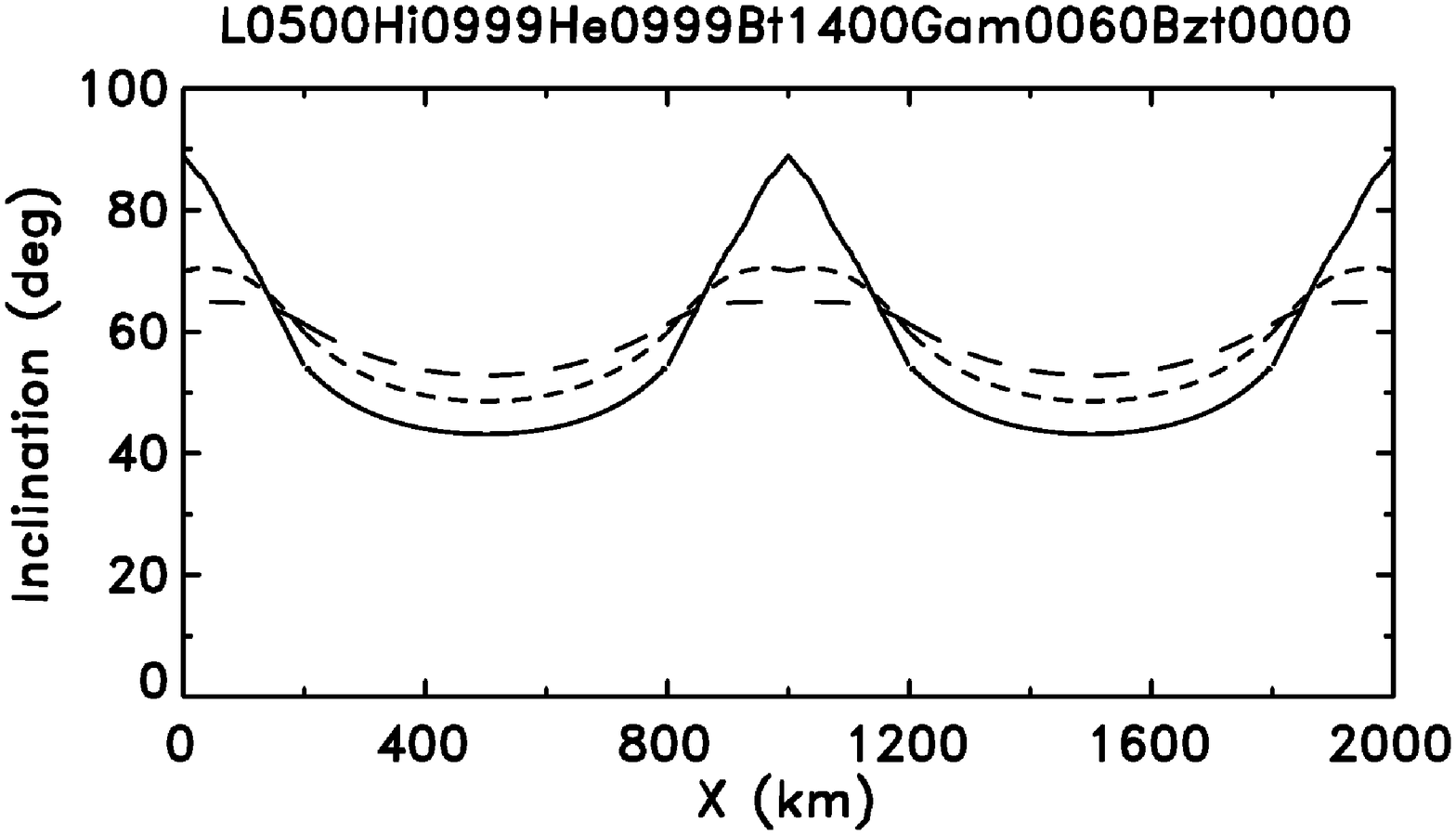}
 \centering \includegraphics[bb=55 112 740 404, clip, width=1.00\hsize]{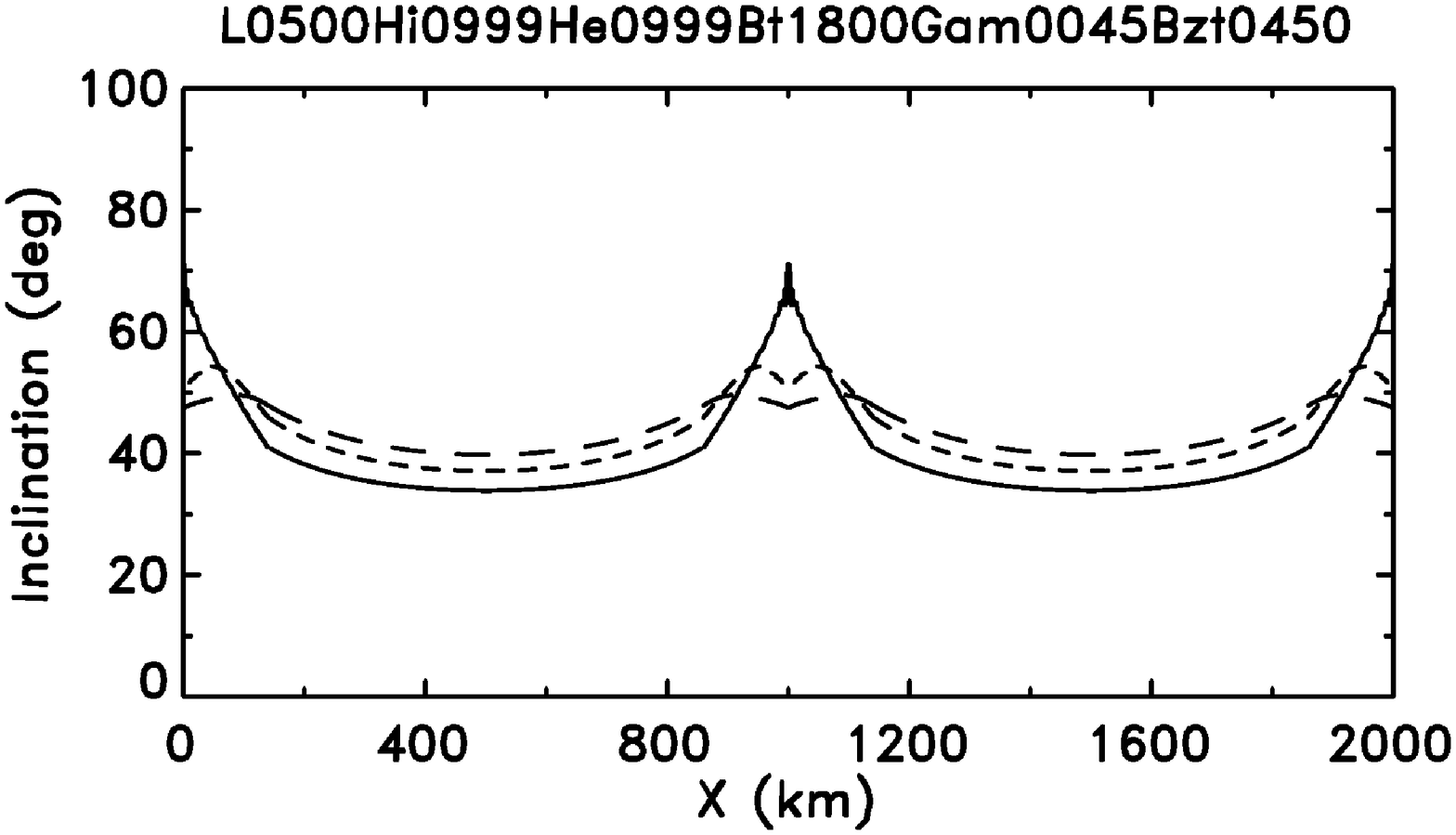}
 \centering \includegraphics[bb=55 68 740 404, clip, width=1.00\hsize]{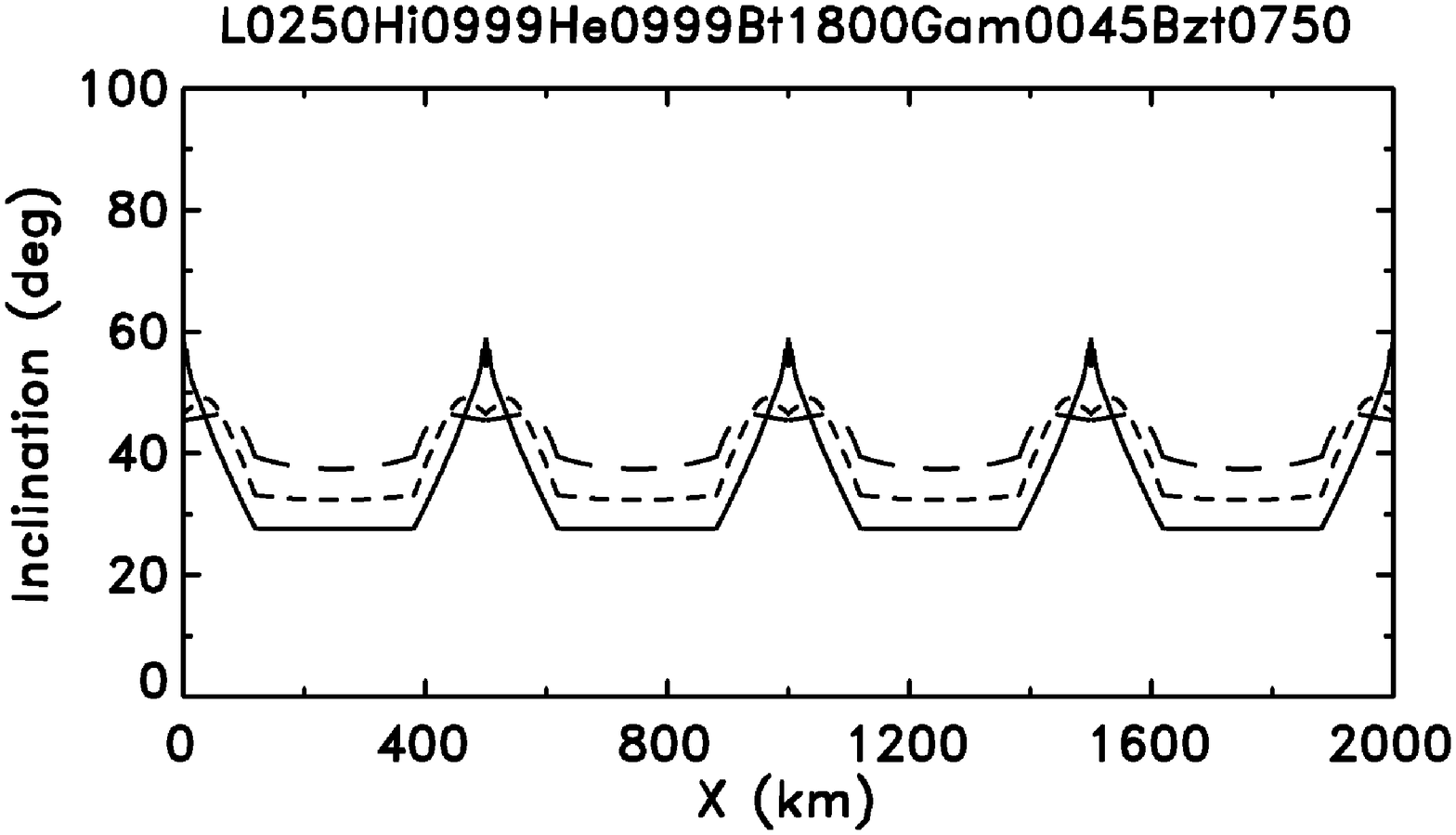}
     \caption{\small The variation of the magnetic field inclination (from the vertical) 
      with horizontal coordinate $x$ at the penumbral photosphere (full) and $z=100$ km (short dashes) and 
      $z=200$ km (long dashes) above the penumbral photosphere. Case I (top), represents the
      outer penumbra, Case II the mid penumbra and Cases III-IV the inner
      penumbra.
             }
     \label{cores1}
\end{figure}
\begin{figure}[htbp]
 \centering \includegraphics[bb=55 112 740 404, clip, width=1.00\hsize]{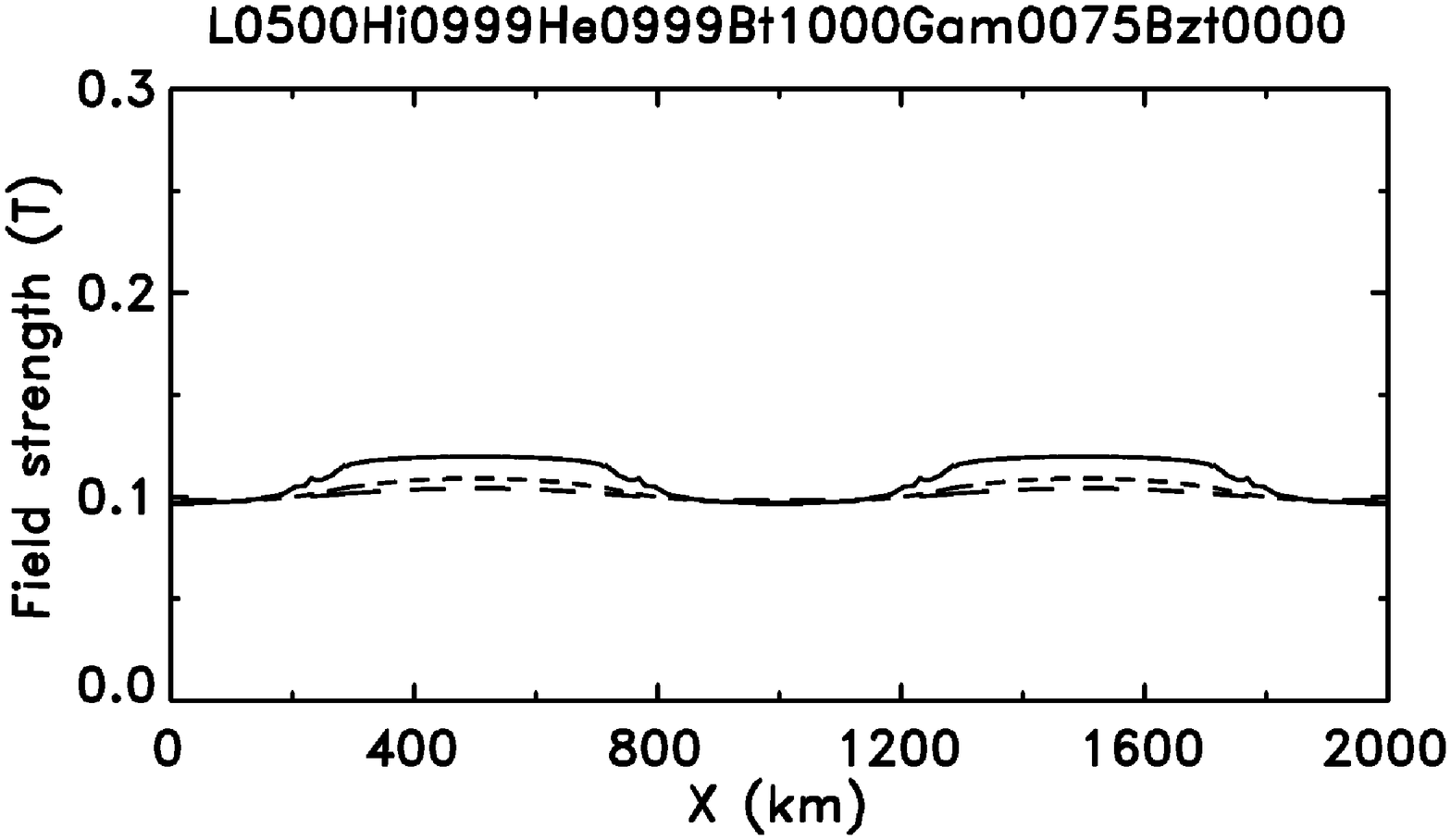}
 \centering \includegraphics[bb=55 112 740 404, clip, width=1.00\hsize]{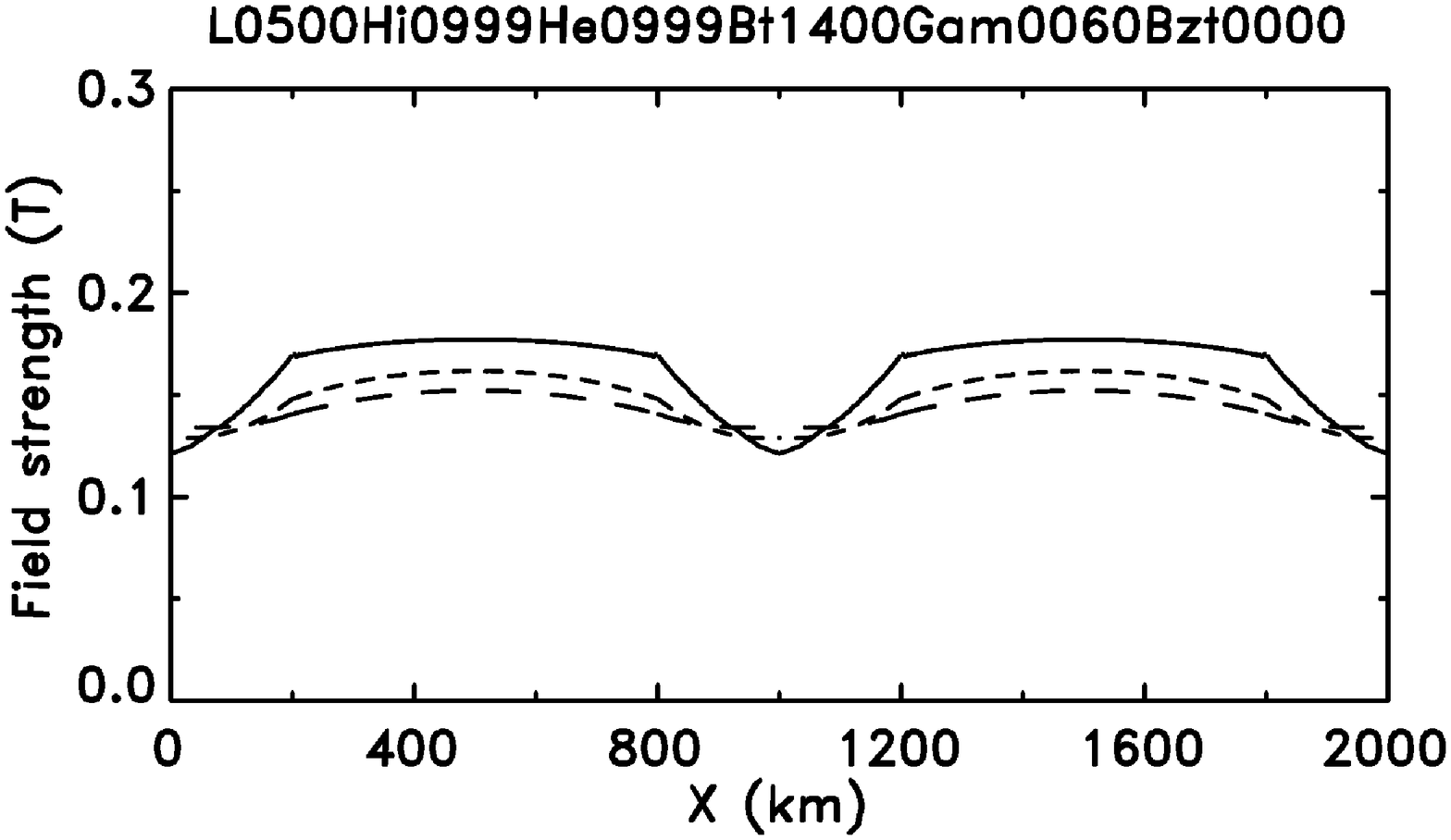}
 \centering \includegraphics[bb=55 112 740 404, clip, width=1.00\hsize]{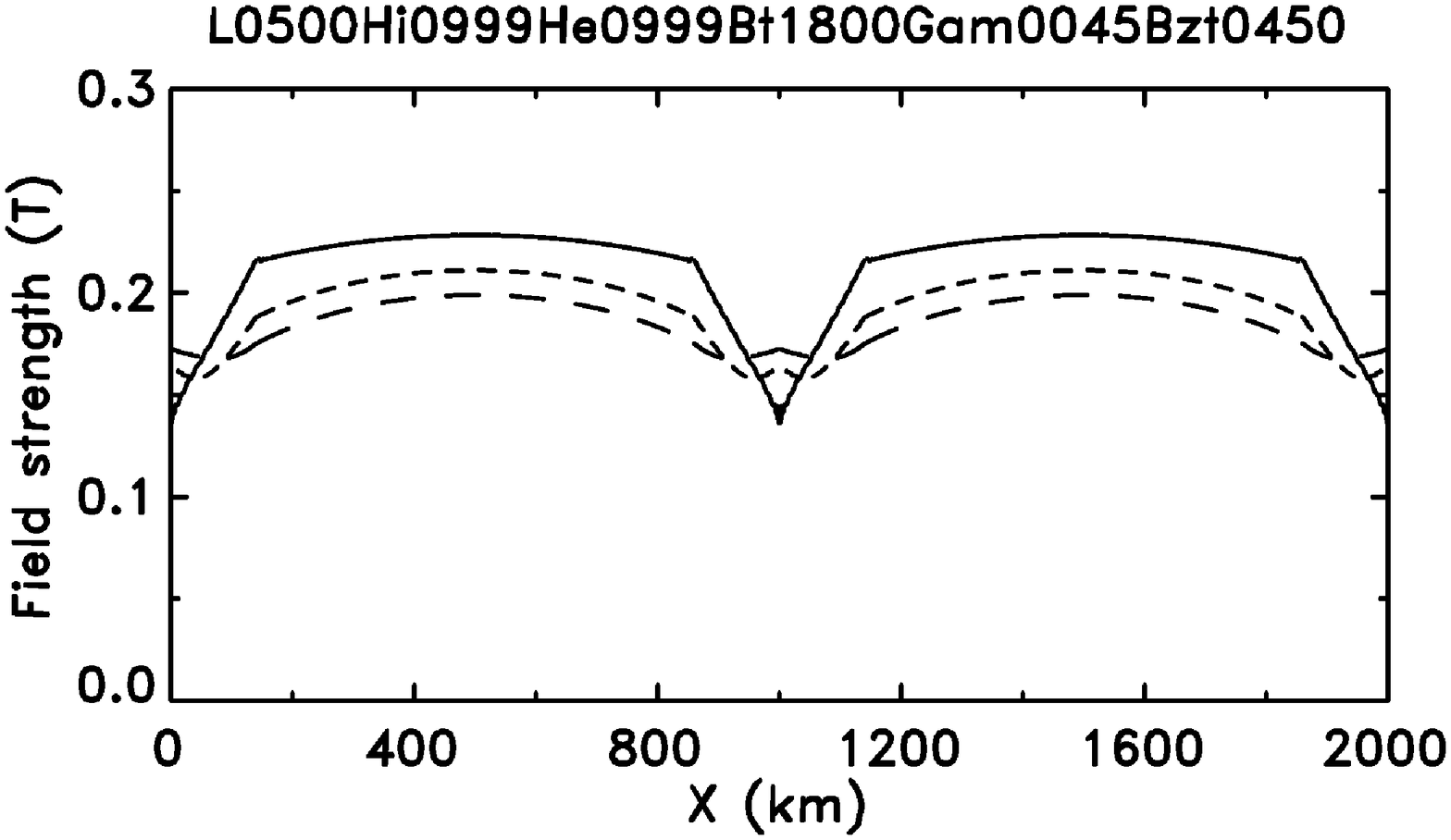}
 \centering \includegraphics[bb=55 68 740 404, clip, width=1.00\hsize]{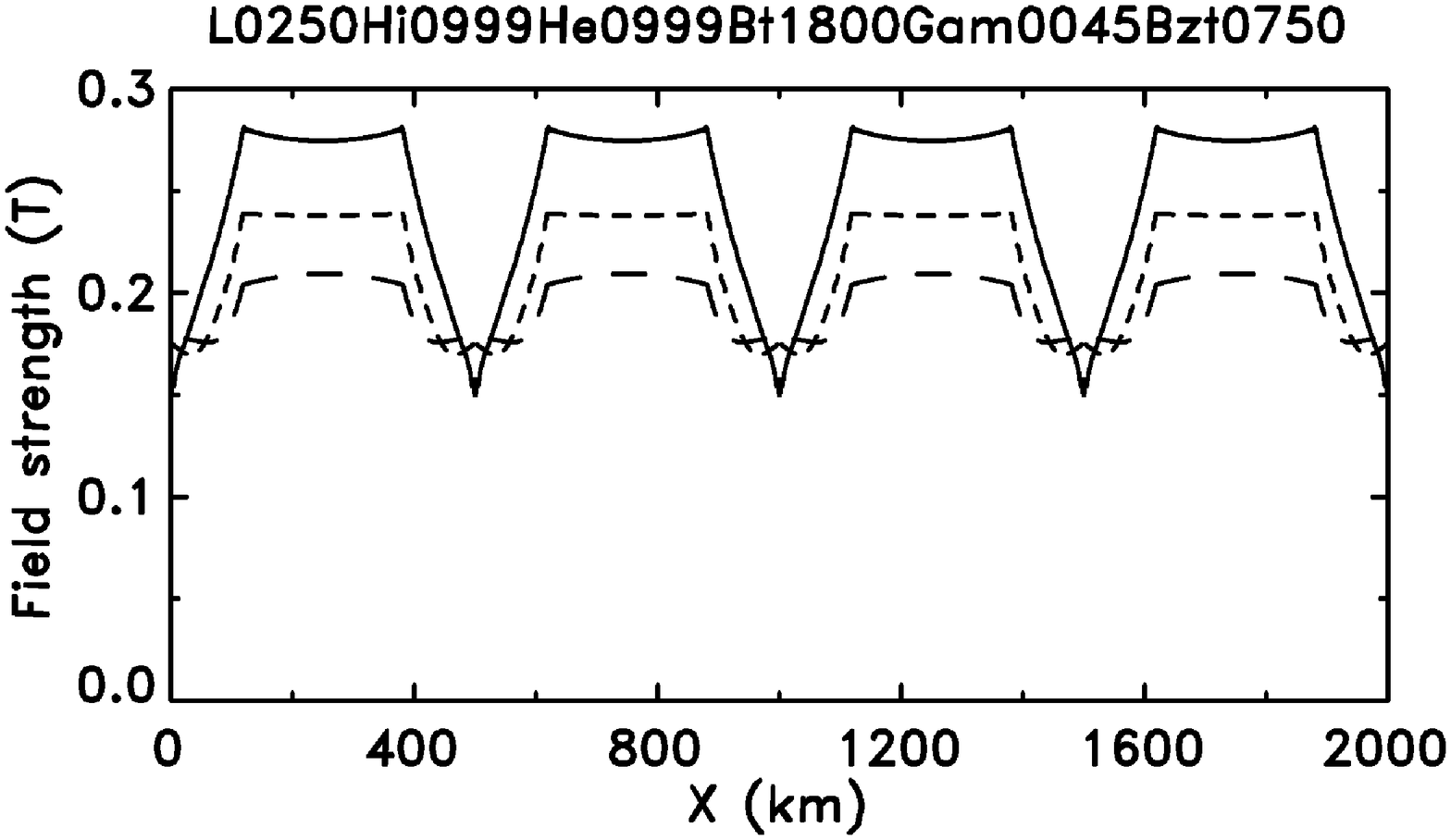}
     \caption{\small The variation of field strength with horizontal coordinate $x$ at the 
        penumbral photosphere (full) and $z=100$ km (short dashes) and
      $z=200$ km (long dashes) above the photosphere. Case I (top), represents the
      outer penumbra, Case II the mid penumbra and Cases III-IV the inner
      penumbra.
             }
     \label{cores1}
\end{figure}

{\it Case II} corresponds roughly to conditions in the mid penumbra. The magnetic field is less 
inclined, $\bar \gamma=60^\circ$, and 
stronger, $\bar B = 140$~mT. Figure 1 shows that the shape of the discontinuity is intermediate 
to that of a flat top and in the form of a {\it cusp}. 
However, for Case II the shape of the 
discontinuity is sufficiently flat that $B_z$ nearly vanishes at the top 
($B_{zg} \approx 2$ mT) leading to a magnetic field that
is nearly horizontal immediately above the center of the gap. Due to the stronger
radial magnetic field component compared to Case I, the Wilson depression is increased
to $130$ km.

{\it Case III} corresponds roughly to conditions in the inner penumbra. The field is even stronger ($180$~mT) and more vertical, $\bar \gamma=45^\circ$, than for Case II. The cusp is now sufficiently pronounced that field lines can easily follow the discontinuity and $B_z$ is therefore non-vanishing, $B_z \approx 44$~mT at the top of the gap. As shown in Fig. 2, the inclination of the magnetic field above the gap is close to $71^\circ$ just above the center of the gap and close to $51^\circ$ at $z=100$ km. The Wilson depression for Case III is larger than for Case II, about $200$ km, due to the increased magnetic pressure above the center of the gap.


{\it Case IV} is identical to Case III except that the separation between two nearby gaps has been reduced to $500$ km. The shape of the discontinuity and the magnetic field topology is similar to that of Case III but compressed by a factor two in the $x$-direction. This leads to a stronger vertical field component and therefore higher field-strength at the top of the gap ($B_z \approx 80$ mT) which reduces the gas pressure in the magnetic component and increases the Wilson depression to $310$ km.

The gradual transformation from a flat-topped boundary (Case I) into a pronounced cusp shaped top (Cases III and IV) can be explained by conservation of magnetic flux. The weak {\it vertical} magnetic field of Case I can be squeezed into a much narrower channel between two field-free gaps, while constrained by magnetostatic equilibrium across the discontinuity, than for Cases III and IV. This allows the top of the gap to extend over a larger horizontal distance and a flat top to form for Case I, but not for Cases III--IV.

No particular significance should be attached to the fact that the limiting value of the inclination is $90^\circ$ for the calculations shown. This is a direct consequence of
our assumption that the field-free gaps are not associated with any Wilson depression. Depending on the (local) radial gradient of that assumed Wilson depression, the limiting inclination will be smaller or larger than  $90^\circ$, thereby also allowing field lines that dip down.


\subsection{Cusps}
As Fig. 1 shows, the gap-tops have a pronounced spike in the inner-penumbra cases: the vertical magnetic field line at the top of the gap `splits in two'. This configuration occurs whenever a field free plasma penetrates into a magnetic field, and is known in the controlled-fusion literature as a `cusp'. Cuspy configurations like the `stellarator' and the `picket fence' play a role as alternatives to the Tokamak configuration, because of their inherent MHD-stability (e.g., Rose \& Clark 1961, Artsymovich, 1964, Haines 1977).

Consider first the simple case $B_y=0$ and, without loss of generality, ignore the gas pressure inside the magnetic region. Let the $x$-coordinate be such that $x=0$ at the cusp point. By symmetry, $B_x=0$ at $x=0$. The gas pressure in the gap is balanced by the magnetic pressure at its boundary, $P_{\rm g}=B^2/2\mu$. This includes, in particular, the cusp point, where $P_{\rm g}=B_z^2/2\mu$. On the axis of the gap and crossing the boundary from the inside to the outside of the gap, $B_z$ thus jumps from zero to a finite value at the cusp point. Measured along the field lines, however, all components of $\bf B$ are smooth, continuous functions. 

Still assuming $B_y=0$, the approximate location $z_{\rm c}$ of the cusp point can be found by balancing the gap pressure $P_{\rm g}(z_{\rm c})$ against the pressure $\bar B^2/2\mu$ of the {\em average} vertical field strength $\bar B_z$ which, unlike the precise value of $B_z$ at the cusp, is known in advance. (This approximation becomes exact in the limit of vanishing gap width). Since the gas pressure does not vanish at any height in the gap, the gap is always terminated by a cusp, as long as $B_y=0$. The situation is more interesting if $B_y\ne 0$. Define a critical height $z_0$, such that $P_{\rm g}(z_0)=B_y^2/2\mu$. The shape of the gap top now depends on the location of $z_0$ relative to $z_{\rm c}$. If $z_0$ is above $z_{\rm c}$ (or $B_y\la\bar B$), the gap has a cusp. In the opposite case, the gap pressure at the gap top can be balanced by $B_y^2/2\mu$. As a consequence, the cusp gradually becomes less pronounced with increasing strength of $B_y$, relative to $\bar B$, and at some value disappears completely. Above this value, $B_z$ vanishes at the top of the gap, and the gap pressure is balanced entirely by $B_y^2$.

\section{Comparison with observations}
Several important characteristics of our simple potential field models are consistent with 
images and magnetograms recorded with the Swedish 1-m Solar Telescope (SST) as well as 
results of two-component inversions by Borrero et. al (2005). Table II summarizes parameters 
calculated from the models and discussed in the following. 

\begin{table}[tbh]
  \centering
  \begin{tabular}{lllllll}
    \hline
    \hline
    Case & $\delta z_{\rm W} $ & $\theta_{cr}  $ & $B_{\rm f}  $ & $B_{\rm m} $ & $\gamma_{\rm f} $ & $ \gamma_{\rm m} $ \\
    & (km) & ($^\circ$) & (mT) & (mT) & ($^\circ$) & ($^\circ$) \\
    \hline
    \hline
    I & 60 & 77 & 100 & 120 & 90 & 54 \\
    II & 130 & 57 & 120  & 180 & 89 & 43 \\
    III & 200 & 35 & 130 & 230 & 71 & 34 \\
    IV & 310 & 21 & 150 & 280 & 58 & 28 \\
    \hline\hline
  \end{tabular}
  \caption{Model properties calculated. $\delta z_{\rm W}$ is the Wilson depression of the magnetic component, $\theta_{cr}$
   the average inclination (from the vertical) of the boundary between the field-free and
   magnetic components, $B_{\rm f}$  is the field strength above 
   the  top of the field-free gap, $B_{\rm m}$ the field strength in the middle of the magnetic  
   component at the  penumbral photosphere and $\gamma_{\rm f}$ and 
   $\gamma_{\rm m}$ are the corresponding inclinations of the magnetic field.} 
  \label{tab:cases}
\end{table}

\subsection{Images and magnetograms}
We first note that the models imply large differences in the {\it appearance} of the field-free and magnetic components at different radial distances in a sunspot. In the outer penumbra, the Wilson depression is small and the field-free gaps occupy a large fractional area close to the surface. For conditions corresponding to the inner penumbra, the Wilson depression is large, on the order $200$--$300$~km, and the field-free gaps occupy a smaller fractional area. {\it The models therefore correspond to field-free gaps that appear as elevated and quite distinct structures in the inner penumbra.} This is consistent with the interpretation (Spruit \& Scharmer 2006) that dark-cored filaments (Scharmer et al. 2002) should be identified with field-free gaps and that the dark cores are located at the center of such gaps. The models are also consistent with the observation that dark cores are easily identifiable in the inner, but not outer, penumbra. The steep "walls" of the field-free gaps, inclined by about $35^\circ$ for Case III and $21^\circ$ for Case IV imply that the limb side "walls" of the field-free gaps cannot be seen at $\pm 90^\circ$ away from the disk center direction at heliocentric distances larger than approximately $35^\circ$ ($\mu = \cos \theta \approx 0.82$). This is consistent with the observation that dark cores are seen with lateral brightenings on {\it both} sides of the dark core at all azimuth angles only for sunspots that are close to disk center (S\"utterlin et al. 2004, Langhans et al. 2005).

The calculated magnetic fields for Cases I--IV have properties that are distinctly different for the inner and outer penumbra and that can be compared to magnetograms. For Case I, corresponding to the outer penumbra, inclination variations are $36^\circ$ but these strong inclination variations are associated only with small, on the order of $20$ mT, fluctuations in the field strength. Over a large fraction of the area, the magnetic field is nearly horizontal. For Cases III and IV, corresponding to the inner penumbra, the magnetic field inclination is in the range of $58$--$71^\circ$ above the field-free gap and 28--34$^\circ$ above the magnetic component close to the  penumbral photosphere. Because of the cusp-shaped magnetic field in the inner penumbra  (Cases III--IV), the inclination of the magnetic field above the gap deviates by about $20^\circ$ from the horizontal plane, in contrast to Case I where a large fraction of the visible surface is associated with a nearly horizontal magnetic field. For Cases III--IV, the difference in field strength close to the penumbral photosphere above the gap and the center of the magnetic component is large, on the order of $0.10$--$0.13$~mT or a factor $5$--$6$ larger than calculated for the outer penumbra. This agrees with magnetograms of sunspots obtained in the wings of the neutral iron line at $630.2$~nm that show strongly reduced magnetic signal in dark cores as compared to the lateral brightenings in the inner penumbra but much smaller variations in field strength in the outer penumbra (Langhans et al. 2005, 2006). The large variations in field strength we find for the inner penumbra are only partly due to horizontal variations in field strength at a fixed height. {\it A major contribution to these variations in field strength is due to a combination of the Wilson depression and field lines converging with depth}. A possibly significant discrepancy between the magnetograms and our magnetostatic penumbra models is that the inclination changes inferred from the magnetograms are small in the inner penumbra, on the order of $10$--$15^\circ$, whereas the models predict nearly two times larger fluctuations. We note that for Case III, these horizontal inclination variations are strongly reduced already $100$ km above the penumbral photosphere, possibly explaining the smaller inclination variations measured from $630.2$nm magnetograms, but also implying smaller fluctuations in field strength from our model than observed.

\subsection{Two-component inversions}
Borrero et al. (2005) have analyzed spectropolarimetric data obtained from neutral iron lines at $1.56~\mu$m and interpreted these data within the context of the uncombed penumbra model (Solanki \& Montavon 1993). In contrast to the \ion{Fe}{I} $630.2$ nm line, these NIR lines are formed within a thin layer close to the photosphere, making a comparison with our models reasonably straightforward. Borrero et al. (2005) pointed out that these lines are, due to their low formation height, insensitive to the location of the upper boundary of the assumed flux tube. Therefore, effectively, their model corresponds to two constant property components. The only exception is in the outer penumbra, where the inversions return a lower boundary for the assumed flux tube that is above the continuum forming layer.

We identify the  `flux tube' component with the atmosphere above the field-free gap, and their background component with our magnetic component.  Compared with the inversions obtained for a sunspot close to sun center (cf. Figs. 5 \& 6 of Borrero et al. 2005), their inversions, as well as those of Bellot Rubio et al. (2004), then show agreement with our model as regards {\it i)} the inferred variation of the field strength with radial distance for both the flux tube and background components, {\it ii)} the inferred variation of the magnetic field inclination with radial distance for both the flux tube and background components and {\it iii)} the variation of the inferred flux tube fill factor with  radial distance in the sense that this fill factor is smaller for the inner than for the outer penumbra.

\subsection{Temperature structure} Our model does not contain an energy equation and we cannot make quantitative predictions about the variation of temperature horizontally and with height. Borrero et al. (2005) found that in the inner penumbra, their flux tubes are hotter than the background atmosphere by about $500$~K but in the outer penumbra  cooler than the background atmosphere by a similar amount. We note that over the $250$ km height assumed to correspond to the vertical extent of the flux tube, the temperature in their background atmosphere drops by approximately $1400$ K. Since in the inner penumbra the flux tube has much higher gas pressure and therefore much higher opacity than the background atmosphere, the higher temperature found for the flux tube  relative to the background at the same height may still be associated with a lower radiation temperature (lower intensity) than for the background atmosphere, as found by Martinez Pillet (2000). If so, the inversions can be consistent with our model also as regards inferred temperatures.

\subsection{What about flows?} 
The model for penumbral gaps presented here only addresses their
equilibrium aspects. Flows of various kinds are observed in the penumbra,
and the question arises to what extent the gappy penumbra model can
accommodate such observations. As pointed out in the above, the
configurations found here have horizontal fields directly overlying the
center of the gap, for conditions corresponding to the outer penumbra. At
first sight, this looks good because the outer penumbra is also the region
with the strongest horizontal flows (Evershed flow), thus allowing the
simplest interpretation of horizontal, field-aligned flows.

On closer inspection this interpretation has problems. The field lines
wrapping around the gap are sufficiently horizontal only over a relatively
short distance, before turning up into the atmosphere. Such a field line
cannot support a steady flow because the rapid decrease of density with
height in the atmosphere would require a rapidly diverging flow speed. In SS06 we speculated that, instead, the flow on
such field lines is episodic and patchy. Observed Evershed flow do indeed show localized and time dependent variations but also a stronger, {\it steady} flow component (Shine et al. 1994, Rimmele 1994, Rouppe van der Voort 2003, Rimmele \& Marino 2006).

A second problem is that flows are observed not only in the outer
penumbra, but also in the inner penumbra where the measured field
inclinations with respect to the horizontal appear to be substantial in
all penumbral components (Bellot Rubio et al. 2004, Borrero et al. 2005), 
including the dark cores (Langhans et al. 2006).
It is clear that this is a generic problem (not just in the gappy penumbra
interpretation), since it shows that irrespective of the nature of the observed 
flows, they cannot be at the same time steady and field-aligned (again, because 
of the decreasing gas density with height). Direct evidence for strong flows in dark cores of filaments, strongly suggesting also a significant vertical velocity component, were first reported by Bellot Rubio et al. (2005). Rimmele and Marino (2006), 
elaborating on the moving flux tube model, claim that the observed upflows turn into horizontal outflows within fractions of an arcsec. However, an analysis of the azimuthal variation of the measured line-of-sight velocities for the individually resolved flow channels, needed to support interpretations involving horizontal flux tubes, is not presented. Whereas the cospatial nature of flows and dark cores, noted also by Langhans et al. (2006), is indisputable, it has not been demonstrated that such flows are parallel to {\it the visible surfaces of} the dark cores.

At the moment there appear to exist only poorly developed ideas for the
time dependence and/or inhomogeneity that would be needed to explain the
observations. Rimmele and Marino (2006) propose inhomogeneities involving 
a tangle of flux tubes crossing
over each other at different heights in the atmosphere, and interpret this
as support for the tube model. The objections to this elaboration are the
same as for the original moving tube model (see SS06), but in a more
extreme form. On account of the high Alfv\'en speeds in the atmosphere,
for example, an inclusion like a tube embedded at an angle with respect to
its surroundings is not in equilibrium, and will change on the Alfv\'en
crossing time over the width of the tube (seconds, for a diameter of
100km). The lack of equilibrium will be even more serious when the tube is
replaced by a tangle of narrower ones.

The gap model, however, opens opportunities that have not been considered
before. On the one hand, the gap contains a convective flow much like
granulation: up in the middle and down on the radiating sides of the gap.
It also allows for (but does not require yet in the present theory)
horizontal flows along the length of the gap. These could be driven by the
variation of physical conditions from the inner penumbra to the edge of
the spot. The moat flow, seen appearing from under the penumbra at its
edge, would already be present in the gaps, for example. Both kinds of
flow should have an effect in Doppler measurements, since the surface of
the gap is so close to $\tau=1$ that the spectral lines used are partly
formed in the field-free region. Recent 2D Spectrometric data  
in the non-magnetic \ion{Fe}{I} $557.6$~nm line at 0.5 arcsec resolution indeed suggest that the Evershed flow peaks close to the photosphere (Bellot Rubio et al. 2006).

\subsection{Stokes spectra} A crucial test for any penumbra model is its ability to reproduce observed polarized spectra. Of particular importance is to reproduce  asymmetric Stokes-V profiles responsible for production of net circular polarization at locations where the magnetic field vector is at large angle with respect to the line-of-sight.  Such strong gradients, in combination with gradients in velocity, are needed to reproduce observed spectropolarimetric data (e.g. Sanchez Almeida \& Lites 2002, Solanki \& Montavon 1993,  Martinez Pillet 2000). The model presented here predicts strong gradients in both the inclination and azimuth angle along the photosphere as well as in the vertical directions. A more direct test must await detailed radiative transfer calculations and flow models.

We note that our conceptually simple model has a more intricate magnetic field configuration than  implemented in inversion techniques based on e.g., the embedded flux tube model. As shown in Fig.~2, the inclination decreases with height (becomes more vertical with height) above the field-free gap, but above the center of the magnetic component, the inclination increases with height. As shown in Fig.~3, the field strength increases with height above the field-free gap and decreases with height above the magnetic component. We speculate that embedded flux tube inversions may respond to such magnetic field gradients by returning a lower boundary for the flux tube (identified with the field-free gap) that is located slightly above the photosphere, as found by Borrero et al. (2006a) for the mid and outer penumbra.

\section{Limitations of the present model}

The model, focusing on the gas pressures and the magnetic field configuration, is coarse in terms of the temperature distribution. A self-consistent temperature structure requires a model of the convection within the field-free gap, and radiative transfer to estimate the radiative flux and cooling near the surface as well as heating of the magnetic component. 
A significant uncertainty is the thermodynamic state of the magnetic flux bundles between the gaps, and the processes that determine their field strength. Another complication is that the field lines are concave towards the field-free gas (much more so in the outer than in the inner penumbra) and therefore the configurations modeled in this paper should in principle be subject to fluting instabilities. We repeat our speculation (SS06) that this interaction may have something to do with generating the Evershed flow. Ultimately, 3D MHD simulations will be required to understand these complicated interactions. 

Our 2D model is such that $B_x$ and $B_z$ are assumed 
divergence-free, thereby implying that $\partial {B_y}/\partial{y}$ must be zero, excluding 
gradients of $B_y$ in the radial direction. This is a consequence of using independent 2D 
models for the inner, mid and outer penumbra. A 3D model would remove this 
inconsistency but does not constitute a problem as regards our main conclusions, 
summarized below.
 
\section{Conclusions and discussion}
We have shown that magnetostatic penumbra models characterized by field-free gaps can be constructed by allowing the shape of the discontinuity to adapt itself to the required force balance between the field-free and magnetic components. This is in contrast to the embedded flux tube model where the geometry, a round flux tube with internal field lines aligned with the flux tube and external field lines wrapping around the flux tube, is assumed given a priori. As pointed out by SS06 and further discussed in Sect. 2, such flux tubes cannot be in magnetostatic equilibrium.

The magnetostatic gappy penumbra model presented here is conceptually simple. We have assumed a potential field interlaced by field-free gaps, a variation of gas pressure with height that is similar to that of the quiet sun for the field-free gap and a gas pressure in the magnetic component that is simply that of the field-free gap scaled by a constant. Further input quantities of the models are the average field strengths and inclinations typical of the outer, mid and inner penumbra. With these constraints, we have calculated boundary shapes and magnetic field configurations.

The calculated models have properties that are distinctly different in the inner and outer penumbra and that agree with observed images and magnetograms. In particular, we find that field-free gaps in the {\it inner} penumbra are cusp-shaped and associated with a magnetic field that is inclined by about $70^\circ$ from the vertical for filaments that are separated by $1000$ km. Here, the magnetic component is associated with a Wilson depression on the order $200-300$ km relative to the field-free component that makes field-free gaps appear as elevated, distinct features. This large Wilson depression, in combination with field lines converging with depth, explains the large variations in field strength inferred from magnetograms and two-component inversions. The steep walls of the field-free gaps explain why dark-cored penumbral filaments are seen with lateral brightenings on both sides of the dark core at all azimuth angles only for sunspots that are close to disk center.
In the {\it outer} penumbra, we find that field-free gaps are associated with flat-topped boundaries and a horizontal magnetic field above the center of the gap. Near the surface, this magnetic field shows large inclination variations horizontally, but only small fluctuations in field strength, in agreement with observations.

We associate the atmospheres above our field-free gap and magnetic component with the flux tube and background components respectively in the inversions of Bellot Rubio et al. (2004) and Borrero et al. (2005). Our models are then consistent with these inversions as regards the variation of field strength and inclination with distance from the umbra for the two components. Our  models also show a widening of the cusp and gradually reduced Wilson depression, consistent with the gradual widening and fading of dark cores, towards the outer penumbra. This is  also consistent with a systematic increase of the flux tube filling factor towards the outer penumbra. 

Whereas our calculations were made by assuming potential magnetic fields, the differences between the inner and outer penumbra are fundamentally due to magnetic flux conservation, constrained by magnetostatic equilibrium and the average properties of the magnetic field assumed. Flux conservation and a strong {\it vertical} field in the inner penumbra forces the field-free gap to narrow and the magnetic component to widen, as compared to what is the case in the outer penumbra, and a cusp to form. These {\it qualitative} differences between the inner and outer penumbra constitute solid results, not likely to change with more accurate models. More realistic magnetic field configurations, intended for detailed comparisons with observations, may however need to take into account the effects of horizontal temperature gradients, convection and flows.

The interpretation of line profiles and polarimetry is frequently formulated in terms of an embedded flux tube paradigm. The practical implementations of such models, however, are generic 2-component inversions, often not particularly consistent with the physics of embedding of a flux tube in a background magnetic field (cf. discussion in section 2). While these inversions therefore cannot be used as support for their embedded tubes, they have allowed general conclusions about magnetic field strength and inclination variations in the penumbra to be drawn from spectropolarimetric data. Since its introduction 13 years ago by Solanki and Montavon (1993),  no magnetostatic embedded flux tube model has yet emerged, and there are good physical reasons why a realistic model is unlikely to materialize. The embedded flux tube model thus remains a conceptual cartoon, useful in a restricted sense as a two-component model for quantifying variations of the magnetic field, temperature, line of sight velocity and other properties within the resolution element.

The moving tube model of Schlichenmaier (1998a, b) leads to a number of predictions that have successfully been tested against observations. In spite of this success, the presence of flux tubes with circular cross sections in the penumbra meets with the same objections as the embedded flux tube models; such flux tubes (if they exist) are more likely to manifest themselves as sheets with an azimuthal thickness much smaller than their vertical extent (Jahn \& Schmidt 1994). Interchange convection in such flux sheets has been suggested as a  possible heating mechanism for the penumbra (Jahn \& Schmidt 1994). However, the long measured lifetimes (on the order of $1$~hr) for filaments (e.g. Sobotka \&  S\"utterlin 2001, Langhans et al. 2005) as well as flow channels (Rimmele \& Marino 2006) leads to the conclusion that this is not a viable heating mechanism (Schlichenmaier \& Solanki 2003). While the flux tubes simulated by Schlichenmaier (1998a,b) appear to explain Evershed flows, such flux tubes or flux sheets therefore also pose severe problems for explanations of penumbral heating. As shown by Schlichenmaier and Solanki (2003), individual flux tubes cannot heat penumbral filaments extending over more than approximately $1000$--$2000$~km. Such flux tubes must either submerge to give room for new flow channels, or heating of the submerged part of the flux tube must occur. Schlichenmaier and Solanki (2003), relying on simulations by Schlichenmaier (2002), speculate that the submerged part of the flux tube is heated radiatively by hotter gas below the photosphere, causing it to reappear as a hot upflow channel. The efficiency of such radiative heating is restricted to shallow depths below the photosphere. Therefore, this explanation ultimately relies on (efficient) convection to provide the needed heat flux. Evidence to support either of the two scenarios discussed above is absent in the highly resolved images analyzed by Rouppe van der Voort et al. (2004). We also note that Rimmele and Marino (2006) found no evidence for downflows along the observed flow channels.

Our model predicts strong vertical and horizontal gradients in both the magnetic field inclination and azimuth angles. In this paper, we have made no attempts to adjust our models to match such gradients with those inferred from inversion techniques. In future work we intend to use our models to calculate synthetic continuum and narrowband images as well as conventional and polarized spectra, using empirically determined temperatures and velocity fields. Our model predicts not only strong magnetic field gradients, but also a strongly warped $\tau = 1$ surface that should enable a number of critical tests with observed high-resolution data.



\acknowledgements{We are grateful to the referee, R. Schlichenmaier, for constructive criticism and suggestions for improvements and to B. Lites and A. Nordlund for comments on an earlier version of the manuscript.}

\end{document}